\documentclass[11pt, a4paper]{article}
\usepackage{jheppub}

\usepackage{amsmath,amssymb}

\newcommand{\tr}{\hbox{tr}}

\newcommand{\ZZ}{\mathbb{Z}}
\newcommand{\id}{\mathbb{I}}
\newcommand{\sgn}{\Sigma}
\makeatletter
\def\@fpheader{\relax}
\makeatother

\title{No Signalling and Unknowable Bohmian Particle
Positions}

\author{Robert C. Helling\\

Ludwig-Maximilians-Universit\"at M\"unchen\\
Theresienstra\ss e 39\\
80333 M\"unchen\\
Germany\\
{\tt E-mail: helling@atdotde.de}}

\abstract{We exhibit in a model with simple dynamics, specifically a
  particle in a square box or two particles in one dimensional boxes, that if an
  experimenter can prepare the initial wave function of a system, the
  maximal information about the positions of Bohmian particles that is
compatible with ``no signalling'' is that they are distributed
according to $|\psi(x)|^2$. In particular, the positions cannot be prepared
independently from the wave function. Any sharper ``actual'' position of the particle must be
inaccessible since it could be used to send signals
instantaneously. This is a consequence of the non-local character of
the Bohmian dynamical law.}

\begin{document}
\maketitle

\section{Introduction}
\setcounter{equation}{0}

Bohmian Mechanics (for an introduction, see \cite{Duerr}) is often presented as an approach to quantum theory
that has an ``ontology'', meaning that it is {\em about} particles
that are characterized by their positions and is in that sense
realistic. The price to pay is that the theory has non-local equations
of motions but is claimed to be observationally indistinguishable
from the textbook version of quantum mechanics (which one could loosely
call Copenhagen style lacking a better name irrespective if one
believes in collapse, many worlds, decoherence or the like).

The Bohmian theory is based on the observation that the quantum
mechanical current
\begin{equation}
  \label{eq:1}
  j = \bar\psi\nabla \psi - (\nabla \bar\psi)\psi
\end{equation}
is conserved for Schr\"odinger type Hamilton operators
\begin{equation}
  \label{eq:2}
  H = -\Delta + V(x).
\end{equation}
From that one defines a velocity field
\begin{equation}
  \label{eq:3}
  v(x) = j/|\psi|^2 = 2\Im(\nabla\psi/\psi)
\end{equation}
and postulates ``particles'' whose position $q(t)$ follow the velocity
field, i.e. 
\begin{equation}
  \label{eq:4}
  \frac d{dt} q = v(q).
\end{equation}
If one starts with a statistical ensemble of such particles with
probability density $|\psi(x)|^2$ at the initial time then at all times
the evolved probability density will be given by $|\psi(x,t)|^2$. Note
well that the wave function $\psi$ evolves according to the time
dependent Schr\"odinger equation and that thus there is no feedback
from the particle postions $q$ to the wave function.

In the Bohmian framework, it is emphasised that all measurements in
the end can be traced back to position measurements (which could be
the position of a pointer on a scale or the position of a black spot
on a photo plate in a double slit experiment) so it is understood that
the $q$ are the positions of the ``real'' or actual particles. They have
objective (realistic, deterministic and in that sense classical) trajectories but
they are quantum in the sense that their dynamical law (\ref{eq:4}) is very
different from Newton's second law. Still, they produce the predictions
of standard quantum mechanics as their probability density traces at
all times the probability density implied by the absolute value
square of the wave function assuming it did at an initial time.

The existence of trajectories appears to be in direct conflict with
standard lore of quantum mechanics as well as the existence of
classical probability densities seems to clash with Bell-type
inequalities. But this tension is usually relieved upon the
realization that as long as one only considers observables that are
functions of the positions only (and not of momenta) these all
commute. And since one deals only with a commutative algebra of
observables, all states can indeed be realized as probability
densities on some classical (configuration) space.

This argument is only true as long as one considers only positions at
one instant of time as in general the positions at different time fail
to commute. This fact, in the context of the Bohm theory, was
emphasised by \cite{CorreggiMorchio, Werner}. In those papers,
positional observables that lead to (violations) of Bell type
inequalities were constructed and here, we strongly build on these
works.

In this note, we will analyse a particular simple example of the
construction in \cite{CorreggiMorchio} and spell out the consequences.

Our model, consisting of a particle in a box, has the advantage of being
elementary solvable so one has full analytic control at all stages as
it comes without devices like double slits, Stern-Gerlach devices or
beam splitters that do not have explicitly know Hamiltonians and in
which technicalities could be hidden or suspected to be hidden. All
observables considered are diagonal in position representation and thus directly
expressible in terms of Bohmian particle positions. The model's
understanding only requires the most elementary quantum mechanics.

\section{The Model}
Often, inequalities of Bell type that demonstrate that Quantum
Mechanics cannot be a local, realistic theory (in the very general meaning of
having a space of states that is a simplex so all states can uniquely
decomposed into extremal states) are expressed in terms of entangled
qubits which are then thought of as realized by spin or helicity
degrees of freedom. For a discussion in the Bohmian context this can
cause problems or confusion since their Hamiltonians are generally
not of Schr\"odinger type (\ref{eq:2}) and the Bohmian trajectories do
not directly apply.

Using Stern-Gerlach type experiments where an
inhomogeneous magnetic field that couples to  a spin degree of freedom
one can translate spin states to positions. But those have the problem that
they cannot be easily accessible to analytic study since they lead to
complicated dynamics.

Of course, the only structural property of a qubit is that it lives in
a two dimensional Hilbert space and has a non-commutative algebra of
operators is acting on it. This can also be realized as a two dimensional
subspace of a positional Hilbert space and this is what we will do in
this note following\cite{CorreggiMorchio, Werner}.

Our system is simply a free particle in a two dimensional square
box, i.e. the Hilbert space is 
\begin{equation}
  \label{eq:5}
  {\cal H} = L^2([-\pi/2,\pi/2]^2)\qquad  \hbox{and}\qquad H = -\Delta
\end{equation}
with Dirichlet boundary conditions. 

At times, it will be convenient to emphasise the bipartite nature of
the two coordinates by the use of an equivalent tensor product language:
\begin{equation}
  \label{eq:6}
  {\cal H} = L^2([-\pi/2,\pi/2]) \otimes L^2([-\pi/2,\pi/2])
\end{equation}
and (very explicitly) a non-interacting time evolution given in terms of
\begin{equation}
  \label{eq:7}
  H = H_1\otimes \id + \id \otimes H_2\qquad \hbox{with}\qquad H_i =
  -\frac{\partial^2}{\partial x_i^2},
\end{equation}
where $\id$ denotes the identity operator. Thus, instead of one
particle in a two dimensional box, this set-up equivalently describes
two independent particles in one dimensional boxes
(i.e. intervals). In the following, we will employ both these one
and two particle interpretations interchangeably. For the two particle
interpretation, it can be beneficial to think of the two intervals as
widely separated to visualize that anything performed on particle one
must not have measurable consequences on particle two as otherwise
no-signalling would be violated. But for the moment, we stick with the
``one particle in a square box'' point of view.

The particle is prepared to
be in the state given by the wave function
\begin{equation}
  \label{eq:8}
  \psi(x_1,x_2) = \frac{\sqrt 2}\pi \left( \cos(x_1)\sin(2x_2)-\sin(2x_2)\cos(x_1)\right).
\end{equation}
With the ground state $\psi_1(x_i) = \sqrt{2/\pi}\cos(x_i)$ of energy
1 and first
excited state $\psi_2(x_i) = \sqrt{2/\pi}\sin(2x_i)$ of $H_i$ of
energy 4, our state can
be written as the entangled state
\begin{equation}
  \label{eq:9}
  \psi = \frac1{\sqrt2}(\psi_1\otimes\psi_2-\psi_2\otimes\psi_1).
\end{equation}
From this form, it is obvious that $\psi$ is an eigenstate of $H$ with
energy 5 and thus stationary. This is also reflected by in the Bohmian
theory where the velocity field vanishes
\begin{equation}
  \label{eq:13}
  v=0
\end{equation}
as the wave function is real and thus according to (\ref{eq:3}), the Bohmian particles do not
move at all.

\begin{figure}
  \centering
  \includegraphics[width=4cm]{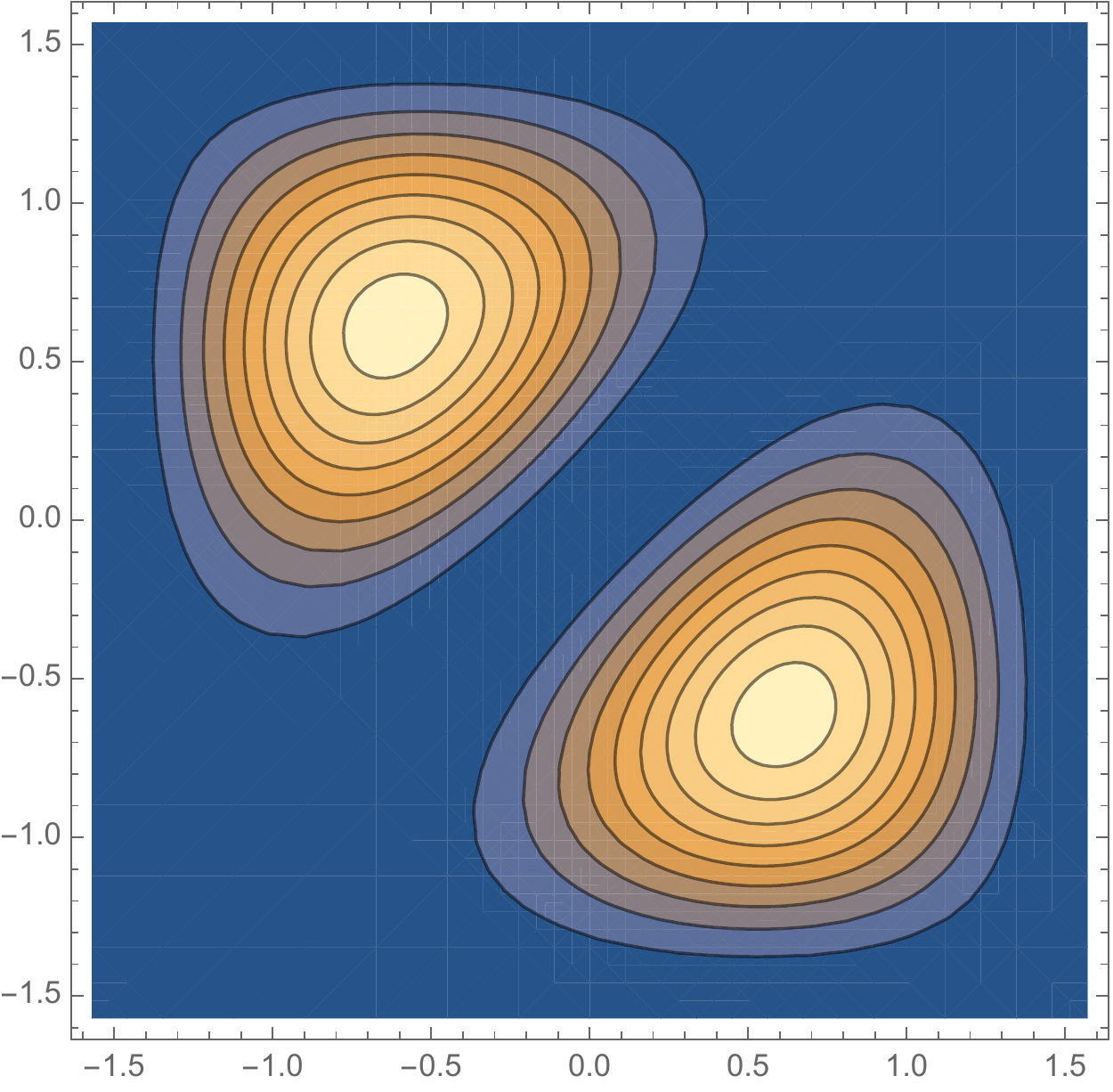}
  \caption{The probability distribution $|\psi(x,y)|^2$}
  \label{fig:prob}
\end{figure}

We will consider two statements about the particle:
\begin{equation}
  \label{eq:10}
  A\colon\hbox{The particle is in the right half of the box}
\end{equation}
and
\begin{equation}
  \label{eq:11}
  B\colon\hbox{The particle is in the upper half of the box}.
\end{equation}
As observables, they can be expressed by the multiplication operators
by
\begin{equation}
  \label{eq:12}
  A = \hbox{sgn}(x_1)=\sgn\otimes\id \qquad B=\hbox{sgn}(x_2)=\id\otimes\sgn,
\end{equation}
where $\sgn$ is the operator that multiplies by the sign function in
position space.
As functions of different coordinates, they commute and are thus
simultaneously observable. As can be observed from
figure~\ref{fig:prob}, by symmetry, both have vanishing expectation
value but their outcome is anti-correlated. One could imagine to do
this experiment by inserting a horizontal or vertical wall in the box
and then detecting the particle in one of the two now separated halves.

We can apply a (Heisenberg picture) time evolution and obtain
\begin{equation}
  \label{eq:14}
  A_t = e^{itH}Ae^{-itH} = (e^{itH_1}\,\sgn\, e^{-itH_1})\otimes\id
\end{equation}
and similarly for $B_t$. As $\psi$ is stationary and $\psi_1$ and
$\psi_2$ are symmetric and anti-symmetric, respectively, one finds
$\langle \psi_i,\sgn\, \psi_i\rangle=0$ and thus
\begin{equation}
  \label{eq:15}
  \langle\psi,A_t\psi\rangle = \langle\psi, B_t\psi\rangle=0.
\end{equation}
As the time evolution does not mix the two tensor factors, the two
observables even commute for different times
\begin{equation}
  \label{eq:16}
  [A_s, B_t]=0
\end{equation}
and thus can be observed without one disturbing the measurement of the
other. 

For the direct application of the Bohmian framework, it is essential
that all these measurements are measurements of positions so that one
can relate the actual observation directly to $q$%
. 
To make use of the quantum nature of entanglement, however,  we need a
non-commutative set of observables which we have in observing $A$ or
$B$ at different times as the Hamiltonian does not commute with $A$ or
$B$ and thus
\begin{equation}
  \label{eq:24}
  [A_s, A_t] \ne 0 \ne [B_s, B_t] \quad \hbox{for } t-s\not\in \pi \ZZ.
\end{equation}
A short calculation yields for their correlation
\begin{equation}
  \label{eq:17}
  \langle\psi, A_sB_t\psi\rangle = -\cos\big[(E_2-E_1)(s-t)\big]
  |\langle\psi_1,\sgn \psi_2\rangle|^2,
\end{equation}
where $\langle\psi_1,\sgn \psi_2\rangle= 8/3\pi$. 
We find that $A$ and $B$ are totally anti-correlated when measured at the
same time but then oscillate to correlation and back.

If one is not happy with the Heisenberg picture of time dependent
observables, one can re-express this time dependence in terms of the
Copenhagen interpretation invoking collapse of the wave function after
the first measurement. One would say that the state is
stationary until the first measurement upon which (depending on the
outcome), the wave function is projected to zero in one half of the
box which results in a state that is no longer an eigenstate of the
total Hamiltonian and thus oscillates which explains the outcome of
the second measurement being an oscillating function of time.

This is the case even though the measurements of $A_s$ and $B_t$ act
in different factors of the Hilbert space and due to their
commutativity do not influence each other.

The spectral projectors for the two eigen-spaces of the operator $A$
are
\begin{equation}
  \label{eq:142}
  P_\pm = \frac 12 (\id \pm \Sigma)\otimes\id= \theta(\pm x)\otimes \id,
\end{equation}
with the Heavyside step function denoted $\theta$.

According to the collapse prescription, after a measurement of this
observables yields the $i$th eigenvalue, the state collapses to
\begin{equation}
  \label{eq:18}
  \psi_c=\frac{P_i\psi}{\|P_i\psi\|}.
\end{equation}
As from there on, we are only interested only in particle (or
coordinate) two, we
trace over the Hilbert space factor of particle one and obtain a
reduced density matrix
\begin{equation}
  \label{eq:19}
  \rho_r = \tr_1|\psi_c\rangle\langle\psi_c|.
\end{equation}
Without the measurement, this reduced density matrix was simply $\frac
12\id_2$ in the $\psi_1,\psi_2$ basis, but as 
\begin{equation}
  \label{eq:20}
  \langle\psi_1|\theta|\psi_2\rangle = \frac 4{3\pi}\ne 0,
\end{equation}
after the measurement contains off-diagonal entries
\begin{equation}
  \label{eq:21}
  \rho_r=
  \begin{pmatrix}
    \frac 12&\frac 4{3\pi}\\
\frac 4{3\pi}&\frac 12
  \end{pmatrix}.
\end{equation}
This density matrix is diagonalized in the basis of eigenvectors
$(\psi_1\pm\psi_2)/\sqrt 2$ with eigenvalues
$\frac12\pm \frac 4{3\pi}= 0.924413$ and $0.0755868$.

\begin{figure}
  \centering
  \includegraphics[width=6cm]{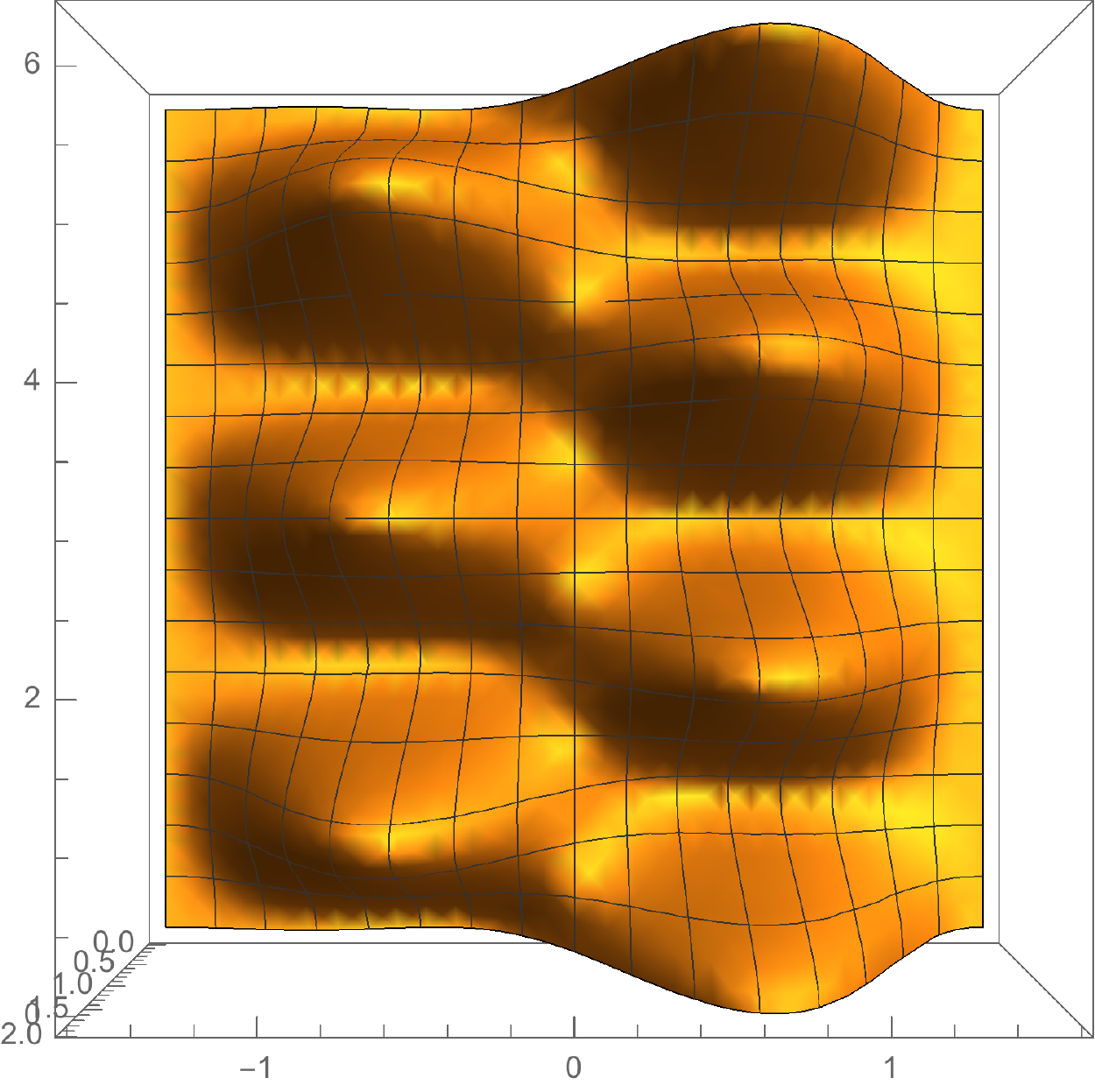}  \includegraphics[width=6cm]{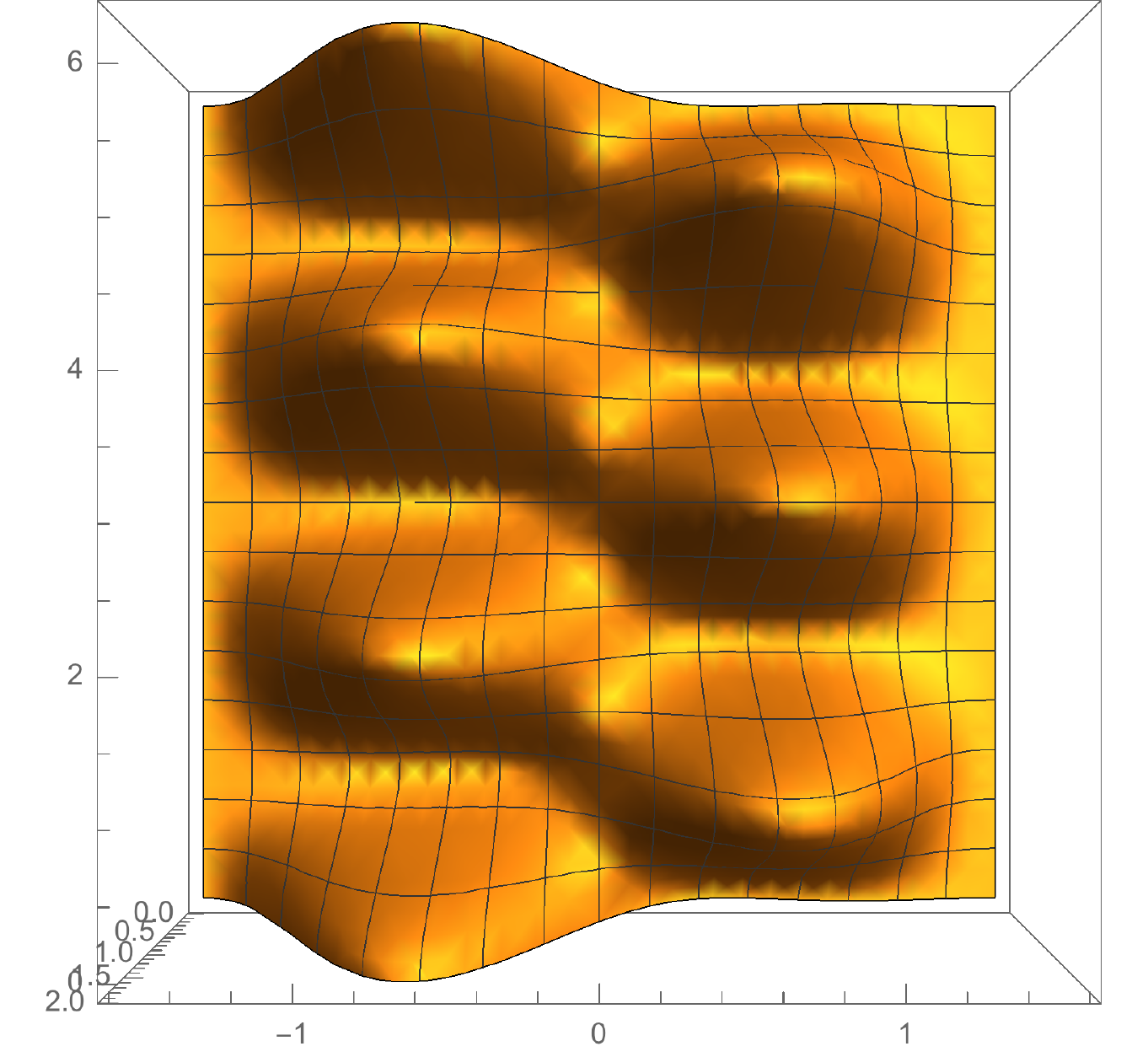}  \includegraphics[width=6cm]{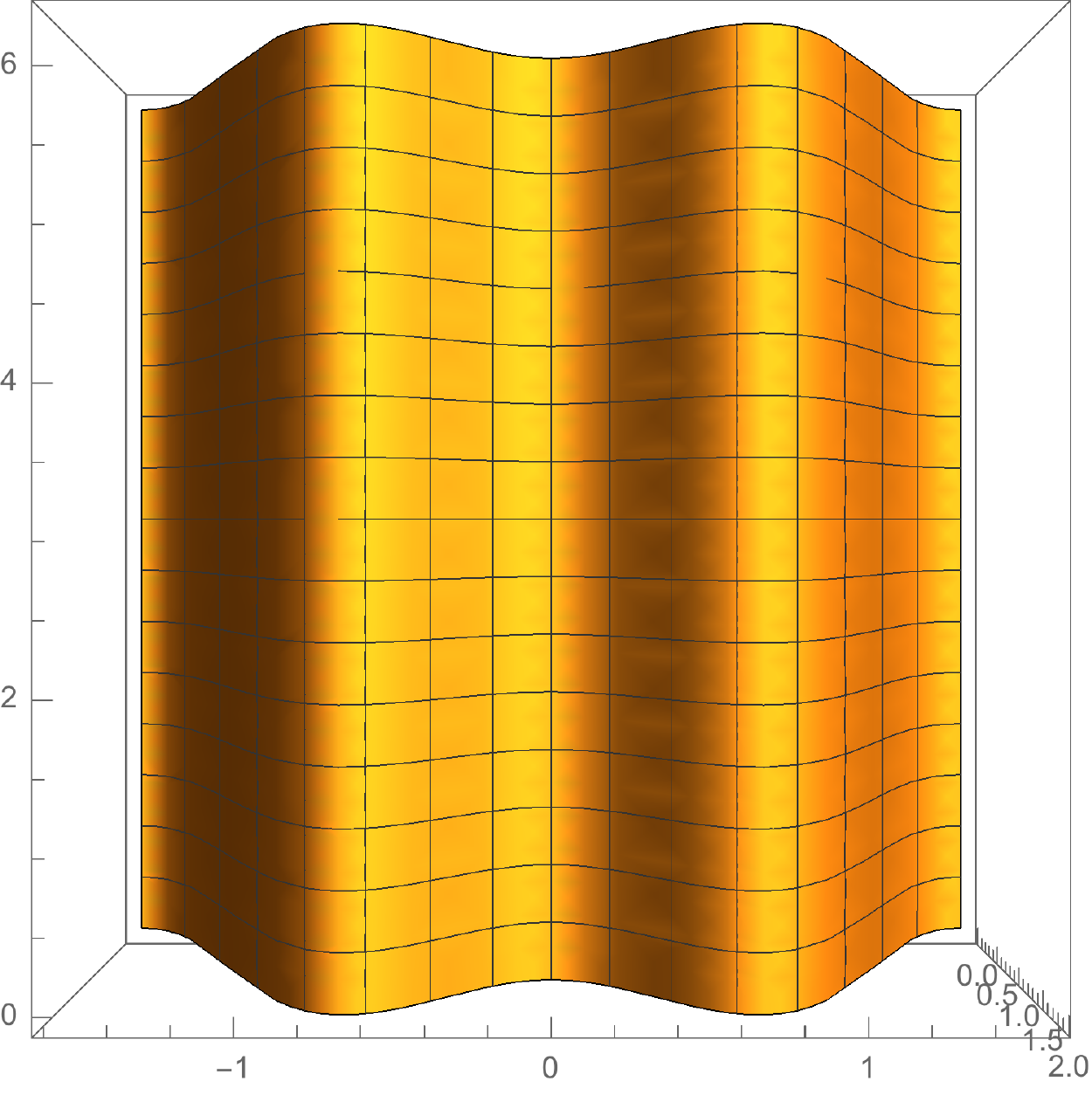}
  \caption{The probability distribution for particle two as a function
    of time (pointing up) after the
    observation of particle one if particle one has been found in
    left/right half (top) and sum of the two lacking information about
  particle one (bottom)}
  \label{fig:collapse}
\end{figure}

Clearly, these eigenvectors are no longer eigenfunctions of the Hamiltonian of
particle 2 and thus this state is no longer stationary.
The probability density for the second particle to be found in a 
specific position is plotted in Figure \ref{fig:collapse} and one can
see the oscillations that lead to the time dependent oscillation
(\ref{eq:17}) of
the particle two between left an right.
This is conditioned on the first particle being found in the left
half of the interval. If it would have been found in the right half,
the probability density would oscillation in the opposite
direction. If the outcome of the measurement on particle one is not
known, both distributions would have to be superimposed (mixed)
yielding the stationary distribution just as without any measurement on
particle one at all. So the distribution of particle two oscillates
only conditioned on the outcome of the measurement of particle one.

\section{The Bohmian perspective}
After this analysis in the Copenhagen language we will perform the
same in the Bohmian framework.
In order for it to cover the oscillating correlation
as a function of the different times of observation it cannot keep the
particle velocity $v$ zero at all times. So, also here does the
measurement of the $x_1$ coordinate (``left'' or ``right'') influence
the velocity of the Bohmian particles in the $x_2$-direction: They
have to oscillate up and down. Even thought there is no explicit
collapse in the Bohmian framework, the measurement is again measuring
the position of particle one (resulting in a moving particle two). We
don't have to know about the detailed workings of the measurement
process to conclude that the particle has to start moving as otherwise
one would not measure the two-time correlation function
eq.~(\ref{eq:17}) that depends on the difference of the two times. If
it were different there would be an immediate observational difference
between the orthodox and the Bohmian approach.

The Bohmian ontology is only about the momentary positions of the
particles so there is no immediate problem. It can be seen 
that this is in fact everything one can know about these
particles. This is clearest in the realization of the model as two
particles moving in two one-dimensional intervals. If one could
observe their velocity, this would violate No Signalling: By observing
the velocity of particle two one would know if a measurement has been
performed on particle one. This is independent of knowing the outcome
of the experiment. So it must be in principle impossible to measure the
velocity, it is not just that so far nobody cared to measure it. The
particles only have a position unlike for example the particles in
classical Hamiltonian mechanics which have both a position and a
momentum.

One could try to measure the velocity by measuring the position twice,
separated by a short time interval. But then one could argue that the
first measurement would necessarily disrupt the particle so much that
the measured velocity is no longer the velocity of the particle before
trying to measure it.

But there is more: It is usually assumed that we don't have any more
specific information about the initial position of the particles
except that it is distributed according to $|\psi|^2$ (this is known
as the ``quantum equilibrium hypothesis'' in the literature). Again, this is
not about our voluntary ignorance. If we had any more specific knowledge about
the particles' initial position, in particular if there were a way to
prepare it while still preparing the initial wave-function to be
$\psi$, we could measure the position of particle two at time $t$ and
thus (possibly probabilistically) determine if it moved since it had
been prepared which would imply that a measurement at particle one had taken
place, possibly at a large space-like distance.

\begin{figure}
  \centering
  \includegraphics[width=8cm]{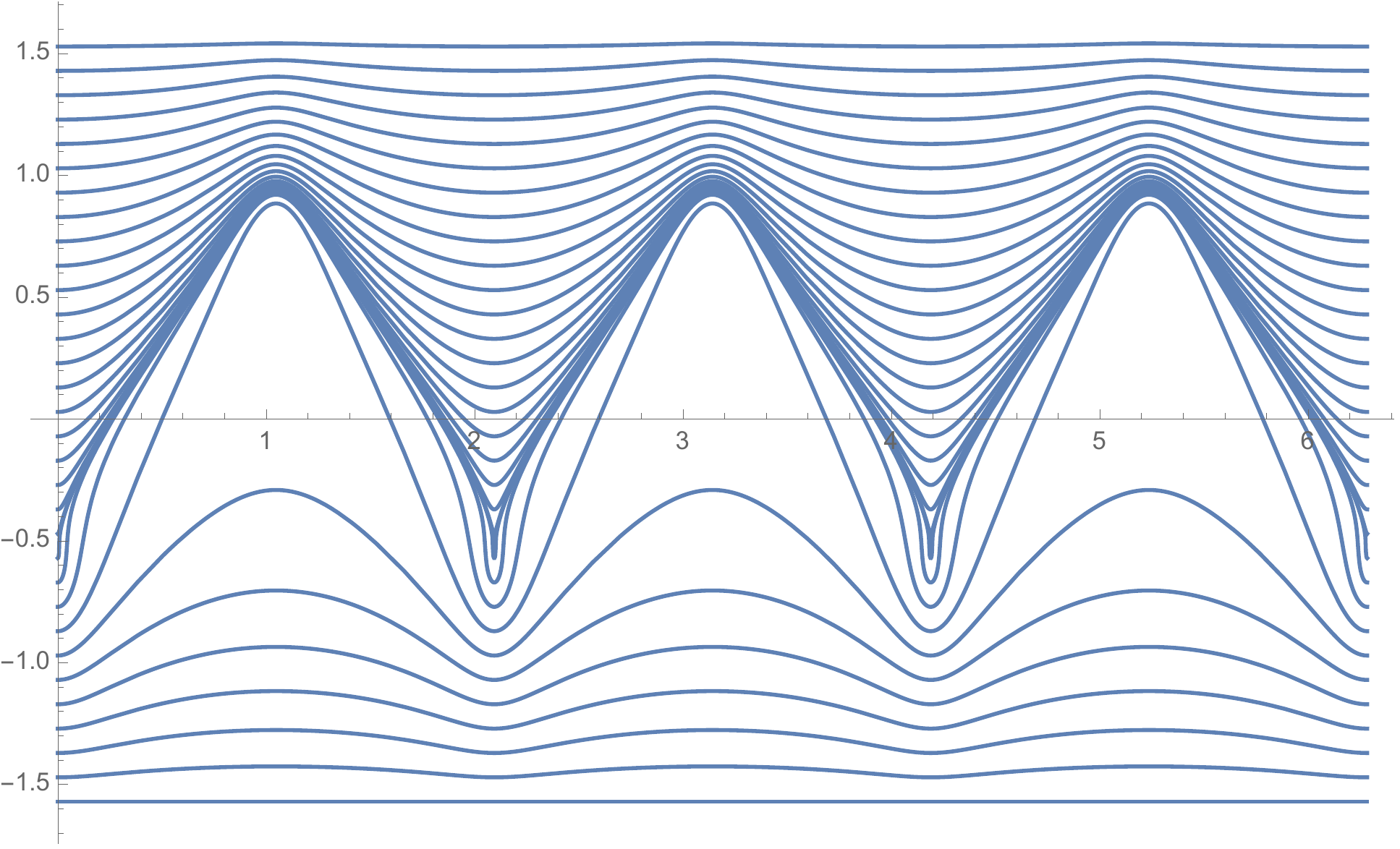}
  \caption{Bohmian trajectories of particle two after particle one has
    been measured in one half of its interval}
  \label{fig:trajectories}
\end{figure}

The oscillating trajectories of the Bohmian particle two after
particle one has been measured (assuming a similar collapse at least
as an effective description of the measurement process) are shown in
Fig.~\ref{fig:trajectories}. Those are key to our claim that if any
information about those particles beyond what is already contained in
$|\psi(x)|^2$ would be accessible to observation, this information
could be used to transmit information instantaneously from particle
one to an observer of particle two (the impossibilitiy of knowing the
distribution of the particles has also been discussed in \cite{Duerr2,valentini1,valentini2}
based on general arguments). 

To this end, let us assume the system of the two particles has been
set up in the quantum state according to (\ref{eq:8}). Let us further
assume that Alice's information about the position of the Bohmian
particle two is described by a probability density $\varrho(x)$. 

In particular, if Alice knew the position of that particle, $\varrho$
would be a $\delta$-function. But her knowledge could be more coarse
grained and be described by a more general probability density. Just
knowing that the quantum state is given by (\ref{eq:8}) is described by
$\varrho(x) $ being equal to $|\psi(x)|^2$, so let us parametrize it
as
\begin{equation}
  \label{eq:22}
  \varrho(x) = \lambda(x) |\psi(x)|^2
\end{equation}

Having any more specific information about the location of the Bohmian
particle would correspond to a non-constant $\lambda$.

Now, Bob wants to transmit one bit of information to Alice. If that
bit is ``0'' he does nothing but if it is ``1'' he observes if particle one
is in the left or the right half (of course without telling Alice the
outcome of his observation).

For Alice waits a short moment of time. If the bit was ``0'' then the
Bohmian velocity was vanishing throughout the experiment. No Bohmian
particle moves and she finds her particle two at a position described
by the original probability density $\varrho$.

If, however, Bob had transmitted a ``1'' and had thus done an observation on
particle one, particle two starts moving with a velocity field (\ref{eq:3})
and the probability density changes due to a non-vanishing divergence
of the current 
\begin{equation}
  \label{eq:23}
  \nabla\left(\lambda|\psi|^22
    \Im\left(\nabla\psi/\psi\right)\right)= (\nabla
  \lambda)\cdot |\psi|^2 v
\end{equation}
which is non-vanishing as long as $\lambda$ is not constant. Thus this
change in position can be detected by observing the position of
particle two which follows a different distribution than $\varrho$.

Only in the case where $\lambda$ is spatially constant, Alice cannot
detect (not even probabilistically) the difference between Bob sending
a 0 and a 1. So we conclude that no-signalling implies that Alice's
knowledge about the Bohmian particles cannot be better than what is
already given by the probability density indicated by the wave
function. The positions of the particles, the additional ingredient of
the Bohmian interpretation, thus must not be knowable (beyond what is
already known in terms of $\psi$) if no-signalling holds.

\section{Signalling and Semi-Classics}
Non-local signalling has of course never been observed in the real
world and if it were it would create immediate problems with causality at
least as long as the world is believed to be realtivistic.

Furthermore, there is no experiment that can distinguish between the
Bohmian interpretation of quantum theory and the ``orthodox''
version. This is because it is possible to take the Bohmian
perspective and then simply ignore the particle positions $q$ to get back
to the orthodox view.

From a structural stand point, this is possible because besides the
equation of motion $(\ref{eq:4})$ for $q$, the particle positions do not
appear in any other equation of motion, in particular they don't
appear in the Schr\"odinger equation that governs the time evolution
of the wave function. There is no feed-back. This is in contrast to
the situation for example in electrodynamics with charged particles:
There, the motion of the particles feels the electro-magnetic forces
due to the field-strength but the motion of the sources also
influences the electro-magnetic field.

If the Bohmian particle postions would source any other field, one
could use an observation of that field as a proxy for the particle
positions and their motions and build a signalling device based on the
set-up described in this note.

There is an attempt to use Bohmian notions as a semi-classical
approximation \cite{Struyve1, Struyve2}: The idea is to treat
particles quantum mechanically but couple those to a classical field
(for example electro-magnetic or gravitational) via $\partial_\mu
F^{\mu\nu}= j_\nu$ with (in our notation)
\begin{equation*}
  j^0(x,t) = \sum_k e\delta(x-q_k(t)),\quad j^i=\sum_k ev_k^i(t)\delta(x-q_k(t)).
\end{equation*}
In the situation of the previous section, particle two is at rest
before the measurement of particle one is performed. It would only
create an electrostatic Coulomb field. But as soon as
the measurement of particle one takes place, particle two starts
oscillating and not only creates a magnetic field but also an
electro-magnetic wave. One could use a radio receiver close to
particle two to detect that a
measurement of particle one has been performed possibly very far away in the
universe.

This shows that for the Bohmian theory to be non-signalling
(and not to get in tension with causality), it is essential that no
observable degree of freedom couples directly to the particle
positions $q$, they have to remain invisible.
 
\section{Conclusion}
In order not to violate no-signalling, one must not be able to know
more about the particle positions than their $|\psi|^2$-distribution,
in particular, it must be impossible to prepare another (possibly
purer) initial state (of knowledge about the position) of the
particles while keeping the preparation of the wave function or just have
more information about the initial state than this particular
probability distribution. In classical physics, the fact that a state
is given by a probability density that is not a single
$\delta$-function peak expresses ignorance about the microscopic
details. 

In a Bohmian world, any further information would immediately lead to the
possibility to send signals faster than the speed of light.

As a consequence, compared to ``orthodox quantum mechanics'', Bohmian
mechanics has further elements of reality (the particle positions)
with a deterministic equation of motion but without the experimenter's
influence on obtaining knowledge of or controlling the initial
conditions further than what is given by the probability distribution
already encoded in the wave function. In this sense, these particles are like
the proverbial angels on the tip of a pin: They exist and but
interaction with them is very limited: If we observe their position we
can have no information where they came from.

This should be contrasted to the Copenhagen interpretation: As we have
explained, the collapse of the wave function is also global and
instantaneous (and thus leads to oscillations of the wave function
conditioned on the outcome of the first measurement). But the wave
function is not directly observable. In this approach it is clear that
causality is only to be imposed at the level of observables. And here
it is clear that locality (and thus causality) hold simply by the
observation that any observable of the form $X\otimes\id$ commutes
with any observable of the form $\id\otimes Y$ and thus actions of Bob
on his system cannot influence anything that Alice experiences in her system.

\acknowledgments I would like to thank various unnamed members of the
``Workgroup Mathematical Foundations of Physics'' as well as Tim
Maudlin and Ward Struyve for helpful
discussions that helped to sharpen the points made in this paper. 

\bibliographystyle{JHEP}
\bibliography{bohmianeangels}

\end{document}